\documentclass[12pt,preprint]{aastex}
\shorttitle{VERITAS Observations of the $\gamma$-ray Binary LS I +61 303}
\shortauthors{G.Maier et al. (VERITAS collaboration)}

\begin{document}

\title{VERITAS Observations of the $\gamma$-ray Binary LS I +61 303}



%

\author{
V.A. Acciari\altaffilmark{20,1},
M. Beilicke\altaffilmark{2},
G. Blaylock\altaffilmark{3},
S.M. Bradbury\altaffilmark{4},
J.H. Buckley\altaffilmark{2},
V. Bugaev\altaffilmark{2},
Y. Butt\altaffilmark{24},
K.L. Byrum\altaffilmark{5},
O. Celik\altaffilmark{6}, 
A. Cesarini\altaffilmark{1,21},
L. Ciupik\altaffilmark{7},
Y.C.K. Chow\altaffilmark{6},
P. Cogan\altaffilmark{12},
P. Colin\altaffilmark{11},
W. Cui\altaffilmark{8},
M.K. Daniel\altaffilmark{4},
C. Duke\altaffilmark{16},
T. Ergin\altaffilmark{3},
A.D. Falcone\altaffilmark{22},
S.J. Fegan\altaffilmark{6},
J.P. Finley\altaffilmark{8},
P. Fortin\altaffilmark{14},
L.F. Fortson\altaffilmark{7},
D. Gall\altaffilmark{8},
K. Gibbs\altaffilmark{1},
G.H. Gillanders\altaffilmark{21},
J. Grube\altaffilmark{4},
R. Guenette\altaffilmark{12},
D. Hanna\altaffilmark{12},
E. Hays\altaffilmark{5,1},
J. Holder\altaffilmark{23},
D. Horan\altaffilmark{5},
S.B. Hughes\altaffilmark{2},
C.M. Hui\altaffilmark{11},
T.B. Humensky\altaffilmark{10},
P. Kaaret\altaffilmark{18},
D.B. Kieda\altaffilmark{11},
J. Kildea\altaffilmark{1},
A. Konopelko\altaffilmark{8},
H. Krawczynski\altaffilmark{2},
F. Krennrich\altaffilmark{9},
M.J. Lang\altaffilmark{21},
S. LeBohec\altaffilmark{11},
K. Lee\altaffilmark{2},
G. Maier\altaffilmark{12,*},
A. McCann\altaffilmark{12},
M. McCutcheon\altaffilmark{12},
J. Millis\altaffilmark{8},
P. Moriarty\altaffilmark{20},
R. Mukherjee\altaffilmark{14},
T. Nagai\altaffilmark{9},
R.A. Ong\altaffilmark{6},
D. Pandel\altaffilmark{18},
J.S. Perkins\altaffilmark{1},
F. Pizlo\altaffilmark{8},
M. Pohl\altaffilmark{9},
J. Quinn\altaffilmark{13},
K. Ragan\altaffilmark{12},
P.T. Reynolds\altaffilmark{19},
H.J. Rose\altaffilmark{4},
M. Schroedter\altaffilmark{9},
G.H. Sembroski\altaffilmark{8},
A.W. Smith\altaffilmark{1,4},
D. Steele \altaffilmark{7},
S.P. Swordy\altaffilmark{10},
J.A. Toner\altaffilmark{1,21},
L. Valcarcel\altaffilmark{12},
V.V. Vassiliev\altaffilmark{6},
R. Wagner\altaffilmark{5},
S.P. Wakely\altaffilmark{10},
J.E. Ward\altaffilmark{13},
T.C. Weekes\altaffilmark{1},
A. Weinstein\altaffilmark{6},
R.J. White\altaffilmark{4},
D.A. Williams\altaffilmark{17},
S.A. Wissel\altaffilmark{10},
M. Wood\altaffilmark{6},
B. Zitzer\altaffilmark{8}
}

\altaffiltext{*}{Corresponding author: gernot.maier@mcgill.ca}
\altaffiltext{1}{Fred Lawrence Whipple Observatory, Harvard-Smithsonian Center for Astrophysics, Amado, AZ 85645, USA}
\altaffiltext{2}{Department of Physics, Washington University, St. Louis, MO 63130, USA}
\altaffiltext{3}{Department of Physics, University of Massachusetts, Amherst, MA 01003-4525, USA}
\altaffiltext{4}{School of Physics and Astronomy, University of Leeds, Leeds LS2 9JT, UK}
\altaffiltext{5}{Argonne National Laboratory, 9700 S. Cass Avenue, Argonne, IL 60439, USA}
\altaffiltext{6}{Department of Physics and Astronomy, University of California, Los Angeles, CA 90095, USA}
\altaffiltext{7}{Astronomy Department, Adler Planetarium and Astronomy Museum, Chicago, IL 60605, USA}
\altaffiltext{8}{Department of Physics, Purdue University, West Lafayette, IN 47907, USA}
\altaffiltext{9}{Department of Physics and Astronomy, Iowa State University, Ames, IA 50011, USA}
\altaffiltext{10}{Enrico Fermi Institute, University of Chicago, Chicago, IL 60637, USA}
\altaffiltext{11}{Physics Department, University of Utah, Salt Lake City, UT 84112, USA}
\altaffiltext{12}{Physics Department, McGill University, Montreal, QC H3A 2T8, Canada}
\altaffiltext{13}{School of Physics, University College Dublin, Belfield, Dublin, Ireland }
\altaffiltext{14}{Department of Physics and Astronomy, Barnard College, Columbia University, NY 10027, USA}
\altaffiltext{16}{Department of Physics, Grinnell College, Grinnell, IA 50112-1690, USA}
\altaffiltext{17}{Santa Cruz Institute for Particle Physics and Department of Physics, University of California, Santa Cruz, CA 95064, USA}
\altaffiltext{18}{Department of Physics and Astronomy, University of Iowa, Van Allen Hall, Iowa City, IA 52242, USA}
\altaffiltext{19}{Department of Applied Physics and Instrumentation, Cork Institute of Technology, Bishopstown, Cork, Ireland}
\altaffiltext{20}{Department of Life and Physical Sciences, Galway-Mayo Institute of Technology, Dublin Road, Galway, Ireland}
\altaffiltext{21}{Physics Department, National University of Ireland, Galway, Ireland}
\altaffiltext{22}{Department of Astronomy and Astrophysics, Penn State University, University Park, PA 16802, USA}
\altaffiltext{23}{Department of Physics and Astronomy, Bartol Research Institute, University of Delaware, Newark, DE 19716, USA}
\altaffiltext{24}{Smithsonian Astrophysical Observatory, Cambridge, MA 02138, USA}


\begin{abstract}
\mbox{LS I +61 303} is one of only a few high-mass X-ray binaries
currently detected at high significance in very high energy $\gamma$-rays.
The system
was observed over several orbital cycles 
(between September 2006 and February 2007)
with the \mbox{VERITAS} array of imaging air-Cherenkov telescopes.
A signal of $\gamma$-rays with energies above 300 GeV
is found with a statistical significance of $8.4$ standard deviations.
The detected flux is measured to be strongly variable; 
the maximum flux is found during most orbital cycles at apastron.
The energy spectrum for the period of maximum emission can be characterized
by a power law with a
photon index of $\Gamma=2.40\pm 0.16_{stat} \pm 0.2_{sys}$ and a flux above \mbox{300 GeV}
corresponding to 15-20\% of the flux from the Crab Nebula.
\end{abstract}

\objectname{LS I +61 303}.

\section{Introduction}

The high-mass X-ray binary \mbox{LS I +61 303} consists of a massive
Be-type star surrounded by a dense circumstellar disk
\citep{Hutchings-1981, Gregory-2002b}
and a compact object.
It has recently been detected as a source of very high energy (VHE)
$\gamma$-rays by the MAGIC telescope \citep{Albert-2006} and
confirmed by VERITAS \citep{Maier-2007}.
This increases the number of $\gamma$-ray binaries to three;
the other two are PSR B1259-63 \citep{Aharonian-2005}
and LS 5039 \citep{Aharonian-2006}.
While the $\gamma$-ray emission in \mbox{PSR B1259-63} is powered
by the relativistic wind of the young 48-ms pulsar, the unknown nature
of the compact objects in \mbox{LS I +61 303} and \mbox{LS 5039} 
does not exclude microquasar-type emission models.
These X-ray binary systems provide unique laboratories for studying particle acceleration by providing
detailed information about the time-evolution of the particle spectrum over the orbital period.

Optical and radio observations of \mbox{LS I +61 303}
show that the compact object orbits the massive star every 26.496
days; the elliptical orbit is characterized by a semi-major axis
of only a few stellar radii and an eccentricity of $0.72\pm0.15$
\citep{Casares-2005,Grundstrom-2007}.
The distance to \mbox{LS I +61 303} is approximately 2 kpc \citep{Steele-1998}.
The radio, optical and X-ray emission of the binary has a component which is clearly modulated
at the orbital period
\citep{Taylor-1982,Mendelson-1989, Leahy-2001, Wen-2006, Smith-2007},
although considerable variation in the lightcurve
is observed from orbit to orbit (e.g.~\citet{Sidoli-2006}).
Orbital modulation in low-energy ($>$100 MeV) or high-energy $\gamma$-rays ($>$300 GeV) is not yet
confirmed \citep{Massi-2004, Albert-2006}.
An association of \mbox{LS I +61 303} with the COS-B $\gamma$-ray source 2CG 135+01 \citep{Swanenburg-1981}
was proposed early on \citep{Taylor-1978},
and observations with EGRET (3EG J0241+6103) for photon energies $>$100 MeV supported this \citep{Kniffen-1997}.
The MAGIC observations in VHE $\gamma$-rays provided the first
firm connection of the site of the variable very-high energy $\gamma$-ray emission with \mbox{LS I +61 303}.

This paper reports on stereoscopic observations of \mbox{LS I +61 303} with the ground-based $\gamma$-ray observatory VERITAS, at 
energies above 300 GeV.
A publication describing contemporaneous X-ray and TeV observations is in preparation \citep{Smith-2008}.
We adopt orbital parameters as derived from radio data by \citet{Gregory-2002} with an orbital period
of $P=26.4960\pm0.0028$ days and zero orbital phase of $T_{0}=\mathrm{HJD} \ 2443366.775$. 
The periastron takes place at phase 0.23, apastron is at phase 0.73, and inferior and superior 
conjunctions are at phases 0.26 and 0.16 respectively \citep{Casares-2005}.

\section{Observations and analysis}

The observations of \mbox{LS I +61 303} were made
between September 2006 and February 2007 during the construction phase of
\mbox{VERITAS},  an array of four 12-m imaging
Cherenkov telescopes.
\mbox{VERITAS} is located at the basecamp of 
the Fred Lawrence Whipple Observatory in southern Arizona 
(1268 m a.s.l.,~31$^{\mathrm{o}}40'30''$N, $110^{\mathrm{o}}57'07''$W, \citet{Weekes-2002}).
The system combines a large effective area ($> 3 \times 10^{4}$ m$^2$) over a wide energy range (100 GeV to 30 TeV)
with good energy (10-20\%) and angular resolution ($<0.14^{\mathrm{o}}$ on an event-by-event basis).

The overall design of all four \mbox{VERITAS} telescopes is identical.
Each telescope employs a 12\,m-diameter tessellated mirror of Davies-Cotton design \citep{Davies-1957}
with 12\,m focal length, mounted on a altitude-over-azimuth positioner.
The reflector comprises 350 hexagonal mirror facets giving a total mirror area of 106\,m$^2$.
The focal plane is equipped with a 499-element photomultiplier-tube (PMT) imaging camera.
The angular pixel spacing is 0.15$^{\mathrm{o}}$, giving a field of view of 3.5$^{\mathrm{o}}$.
Light cones installed in front of the cameras increase the photon collection efficiency and shield
the PMTs from ambient light.
The three-level trigger system of the VERITAS array
allows a substantial suppression of background events at the trigger level.
It especially suppresses events due to local muons which are a considerable background in single-telescope operation.
The first level of the trigger system consists of custom built constant fraction discriminators (CFD), one for each PMT.
All observations described here were made with a CFD threshold of 50 mV, corresponding to approximately 4-5 
photoelectrons.
The second level, a pattern trigger, requires at least three adjacent triggered pixels in order to generate a camera trigger.
The array trigger determines if level-two triggers from individual telescopes are consistent with an 
air shower. 
A coincidence of at least two telescopes triggering within a time window of 100 ns is required.
When the array is triggered, PMT signals in each telescope
are digitized using \mbox{500 Megasample/second} custom-built flash-ADC electronics.
To achieve a large dynamic range, an autoranging gain switch extends the dynamic range from 256 to 1500.
In the event of a telescope trigger, all signals (i.e.~a 48 ns long flash-ADC trace) 
from all channels in all telescopes are read out as a stereo event.
A description of the technical details and the performance of VERITAS can be found in 
\citep{Holder-2006, Maier-2007b} and references therein.

Observations of \mbox{LS I +61 303} with VERITAS were made with different configurations as the array construction proceeded; no changes to
the individual telescopes were implemented in this period.
Observations from September 2006 to November 2006 were made with a two-telescope system while those
for January and February 2007 were made with three telescopes.
The typical array trigger rate was 90 events/s
for the two-telescope system (with a dead time of 4\%), and
160 events/s (7\% dead time) for the three-telescope system.
Data were taken on moonless nights in ``wobble-mode'', wherein the source was positioned at a fixed offset from
the camera center.
This has the advantage that no off-source observations are necessary for background estimation.
The conservative energy threshold after the analysis cuts adopted here is
300 GeV at $30^{\mathrm{o}}$ 
zenith angle.
The dataset consists of 45.9 hours of observations after quality cuts and dead-time corrections. 
The quality cuts remove all data taken under non-optimal conditions (e.g.~bad weather, runs with large trigger-rate
fluctuations, runs with malfunction of the telescope system).
There is good coverage of \mbox{LS I +61 303} for orbital phases from 0.4 to 0.9, but since  
observations were precluded by the bright moon, no observations were possible during orbital phases from 0.9 to 0.2.
Table \ref{tab:observations} shows the most important observational parameters for each of the five observation periods.

The first step in the analysis is the calibration and integration of the flash-ADC traces, as
described in detail in \citep{Holder-2006}.
The calibration is divided into several sections.
The absolute calibration uses a laser system in order to determine the signal size produced by
single photons.
In the relative calibration, laser events are used to calculate the relative gains of the pixels
and timing differences due to path length differences in the cabling of each channel.
Pedestal events, injected at 1 Hz during an observation run, are used for an estimation of the voltage
offset in each flash-ADC trace and the noise levels due to the night sky background and electronics.
The amount of charge in each flash-ADC trace is calculated by summing the samples for a given window
size.
A two-pass summation method is used here, allowing the integration of the flash-ADC traces with an optimal
signal-to-noise ratio.
The resulting image of an air shower is cleaned in order to remove pixels which contain mainly background light.
The cleaning consists of a two-level filter removing all pixels
with an integrated charge smaller than 5 times their pedestal standard deviation and any pixel that
are adjacent to these higher threshold pixels and having signals smaller than 2.5 times their 
pedestal standard deviation.
The shower image is then parameterized with a second moment analysis \citep{Hillas-1985}.
The direction of origin of the $\gamma$-ray on the sky and the impact parameter of the shower core on the ground are reconstructed
using stereoscopic techniques \citep{Hofmann-1999, Krawczynski-2006}.
At least two images with an integrated charge per image $>$ 400 digital counts ($\approx$75 p.e.)
and a maximum image distance from the center of the camera of less than 1.2$^{\mathrm{o}}$ are required in this reconstruction stage. 
The majority of the far more numerous background events are rejected by comparing the shape (i.e.~width and length)
of the event images in each 
telescope with the expected shapes of $\gamma$-ray showers modeled by Monte Carlo simulations.
These so-called \textit{mean-scaled width} and \textit{mean-scaled length}
parameters \citep{Konopelko-1995, Krawczynski-2006} are calculated with lookup tables based on Monte Carlo simulations.
The lookup tables contain
the median and 90\%-widths of the image parameter width ($w_{MC}$, $\sigma_{width,MC}$) and length ($l_{MC}$, $\sigma_{length,MC}$)
as a function of impact parameter $R$, integrated charge per image $s$, and zenith angle $\Theta$:
\begin{displaymath}
 mscw = \frac{1}{N_{Images}} \left( \sum_{i}^{N_{Images}} \frac {width_i - w_{MC} ( R, s, \Theta ) }{\sigma_{width, MC} ( R, s, \Theta )} \right) 
\end{displaymath}
and similar for mean scaled length.
The cuts applied here are $-1.2 < $ mean scaled width/length $< 0.5$ and 
reconstructed distance of shower core position from the center of the array $<$ \mbox{250 m}.
This and an additional cut on the arrival direction of the incoming $\gamma$-ray ($\Theta^2<$\mbox{$0.025$ deg$^2$})
reject more than 99.9\% of the cosmic-ray background while keeping 45\% of the $\gamma$-rays.
All the cuts are optimized a priori with Monte Carlo simulations of $\gamma$-ray- and hadron-induced air showers.
The background in the source region is estimated from the same field of view
using the ``reflected-region'' model and the ``ring-background'' model
as described in \citet{Berge-2007}. 

The energy of each event in the source and background region is estimated 
from Monte Carlo simulations
assuming that the primary particle is a $\gamma$-ray.
The calculation uses lookup tables and determines the energy of an event
as a function of impact parameter, integrated charge per image, and zenith angle.
Collection areas at different zenith angles for $\gamma$-rays are produced from Monte Carlo simulations 
following \citep{Mohanty-1998}; 
collection areas are interpolated between zenith angle bins to the zenith angle of each event.
The limited energy resolution is taken into account by calculating the collection areas
as a function of reconstructed energy.
The differential flux $dN/dE$ in each energy bin of width $\Delta E$ is then calculated with the number of 
events in the source region ($N_{source}$) and in the background region ($N_{bck}$),
the ratio of the size of the source region to the size of the background region ($\alpha$),
the dead time corrected observation time $T$, and the collection area $A$:
\begin{displaymath}
dN/dE = \left( \sum_{i=1}^{N_{source}} 1/A_{i} - \alpha \sum_{i=1}^{N_{bck}} 1/A_{i} \right) / T / \Delta E
\end{displaymath}
The dependence of the collection area on the spectral index is taken into account by an 
iterative process. In each step of the iteration, collection areas are calculated using the 
spectral index obtained in the previous step. 
The iteration stops when convergence is achieved, usually after 2-3 steps.
In order to accurately calculate source fluxes and energy
spectra, a  model of the telescope response to air showers 
has been developed.
The Monte Carlo simulations take all relevant processes and efficiencies in the
development of the air shower through the atmosphere, the propagation of 
Cherenkov photons through the optical system of the telescopes, and the response of the camera and 
electronics into account; see \citep{Holder-2006} for more details.
The systematic error in the energy estimation is dominated by uncertainties and variabilities
of the atmospheric conditions, in the Monte Carlo simulations, and in the overall photon collection
efficiency.


\section{Results}

\begin{figure}
\plottwo{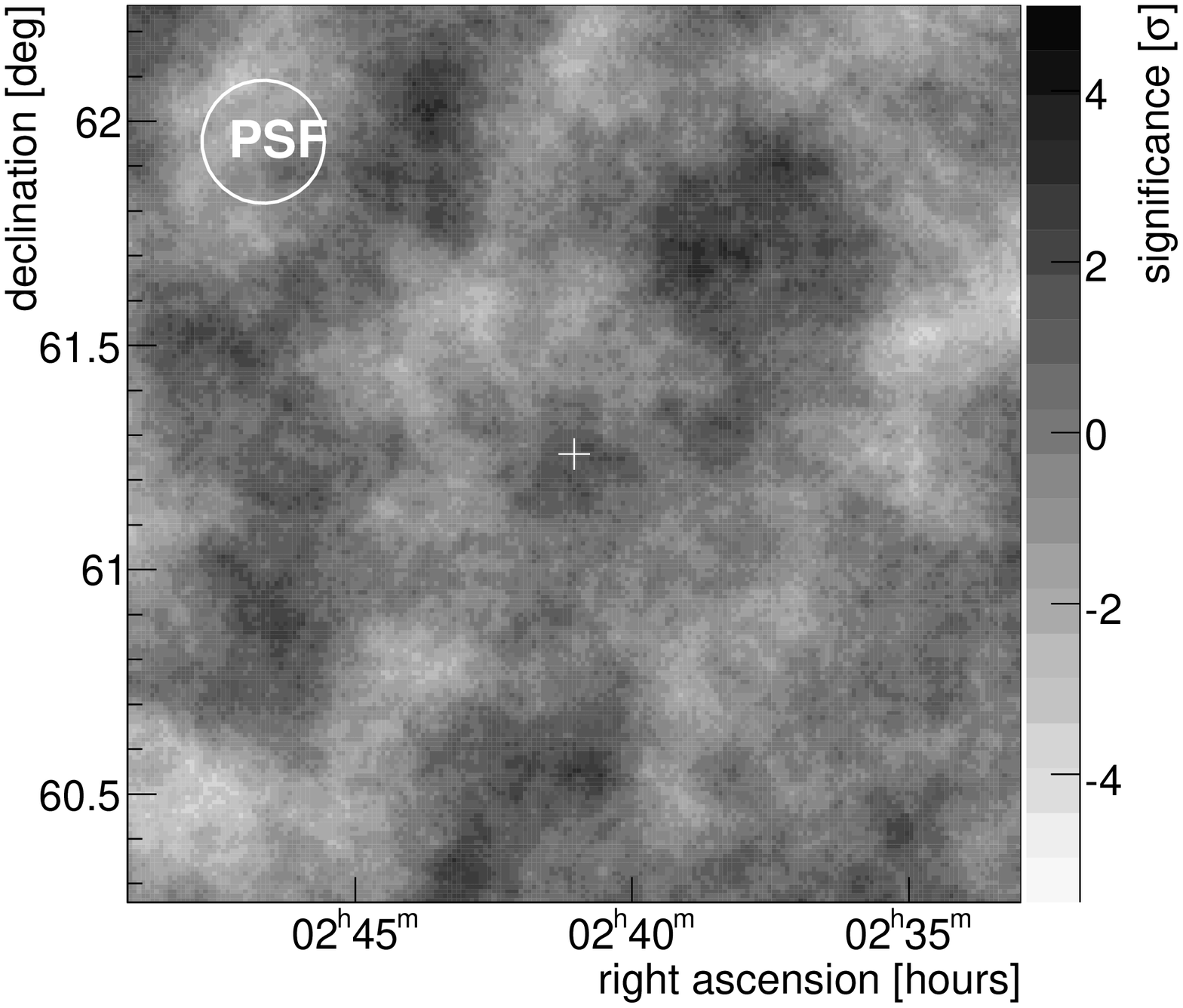}{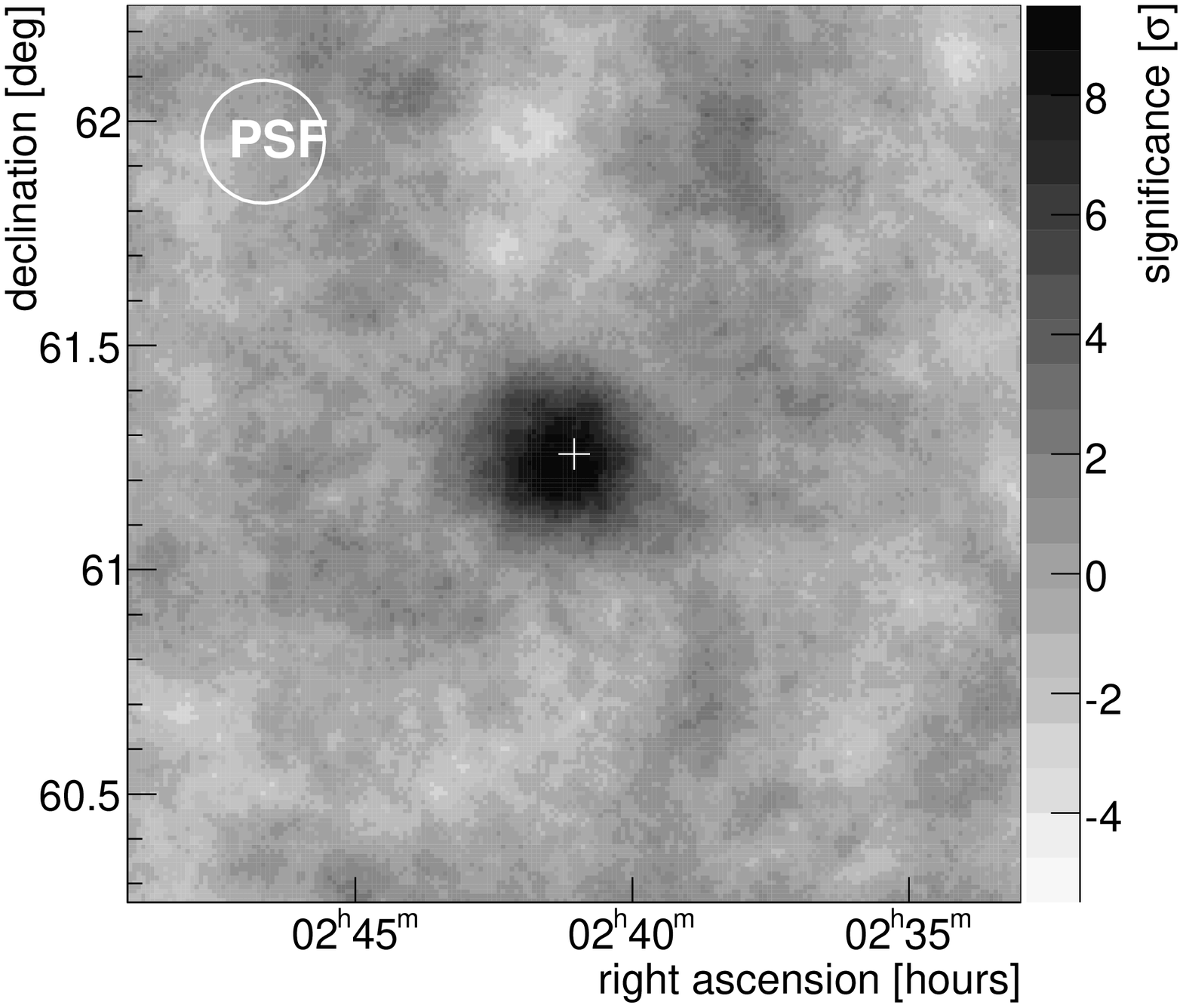}
\caption{\label{fig:SkyPlot} 
Significance map of the region around \mbox{LS I +61 303} in equatorial J2000 coordinates.
Left: Observations during orbital phases 0.8 to 0.5 (18.5 h of data).
Right: Observations during orbital phases 0.5 to 0.8 (around apastron, 25 h of data).
Significances are not corrected for number of trials.
The position of the optical source \citep{Perryman-1997} is indicated by a white cross.
The background is estimated using the ring-background model.
Neighboring bins are correlated.
Significances are calculated using the method of \citet{Li-1983}, equation 17.
}
\end{figure}

We have detected \mbox{LS I +61 303} as a source of $\gamma$-rays with 
energies above 300 GeV, at a total significance of $8.4$ $\sigma$ at the position of the optical counterpart.
The source has been found to be variable, as earlier measurements by MAGIC suggest. 
The probability that the measured fluxes from \mbox{LS I +61 303}
are constant with time or orbital phase has been determined with a $\chi^2$-test to be less than
$\sim10^{-9}$.
The two-dimensional sky map of significances for the region around \mbox{LS I +61 303}
shows a strong detection for orbital phases 0.5-0.8, i.e.,~around apastron 
(Figure \ref{fig:SkyPlot}).
No signal has been found for orbital phases 0.8-0.5.
The position of the peak of the $\gamma$-ray emission, reconstructed by a fit to the uncorrelated map of
excess events with a two-dimensional normal distribution, is in agreement with the position of the 
optical source:
$\Delta_{\mathrm{RA}}=147\pm73_{\mathrm{stat}} \ \mathrm{arcsec},
  \Delta_{\mathrm{dec}}=-34\pm30_{\mathrm{stat}} \ \mathrm{arcsec} $ (corresponding
  to RA$_{\mathrm{J2000}}$ $2^{\mathrm{h}}40^{\mathrm{m}}41^{\mathrm{s}}$, 
Dec$_{\mathrm{J2000}}$ $61^{\mathrm{o}}13'12''$).
The systematic uncertainty is $\pm90$ arcseconds\footnote{
There were no optical pointing monitors installed during these observations with the array under construction.
The pointing accuracy is expected to improve to $<$15 arcsec for the completed system.}.
This is consistent with the position of the TeV excess reported by MAGIC.
The morphology of the excess is compatible with the expected distribution due to a point source.
The VERTIAS source name is VER J0240+612.

\begin{figure}
\plotone{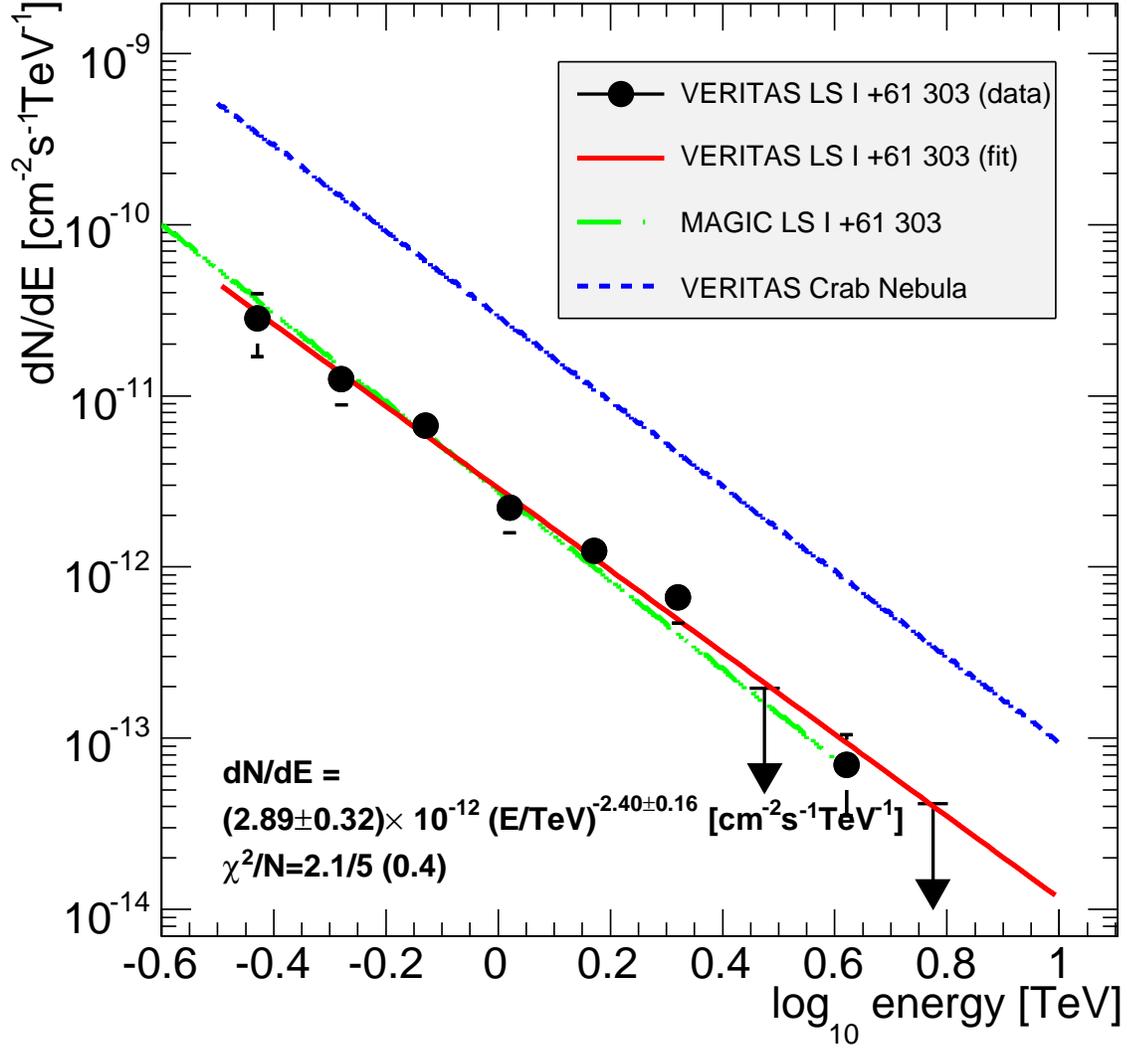}
\caption{\label{fig:EnergySpectrum}
Differential energy spectrum of VHE photons above 300 GeV for \mbox{LS I +61 303}
around apastron (orbital phases 0.5-0.8).
The markers indicate measured data points, the continuous line a fit assuming a
power-law distribution of the data.
Downward pointing arrows indicate upper flux limits (99\% probability, \citet{Helene-1983}) 
for bins with significances below 2$\sigma$.
For comparison, the energy distributions of \mbox{LS I +61 303}
published by the MAGIC collaboration for
orbital phases 0.4-0.7 \citep{Albert-2006} 
and of the Crab Nebula
measured by VERITAS (September to November 2006)
(reconstructed as dN/dE = $2.85\times10^{-11}\cdot \mathrm{E}^{-2.49}$cm$^{-2}$s$^{-1}$TeV$^{-1}$)
are indicated by dotted and dashed lines, respectively.
}
\end{figure}

In Figure \ref{fig:EnergySpectrum} we show the differential energy spectrum during the high-flux phases 0.5-0.8
for $\gamma$-ray energies between 300 GeV and 5 TeV.
The integrated flux of VHE photons in this energy range is $(8.13\pm1.02)\times 10^{-12}$ cm$^{-2}$s$^{-1}$.
The shape is consistent with a power law \mbox{dN/dE = C$\times$(E/1 TeV)$^{-\Gamma}$}
with a photon index $\Gamma=2.40\pm0.19_{stat}\pm 0.2_{sys}$
and a flux normalization constant
$\mathrm{C}=(2.89\pm0.32_{stat}\pm 0.6_{sys})\times 10^{-12}$ cm$^{-2}$s$^{-1}$TeV$^{-1}$.
The $\chi^2$ of the fit is 2.1 for 5 degrees of freedom.

\begin{figure}
\plotone{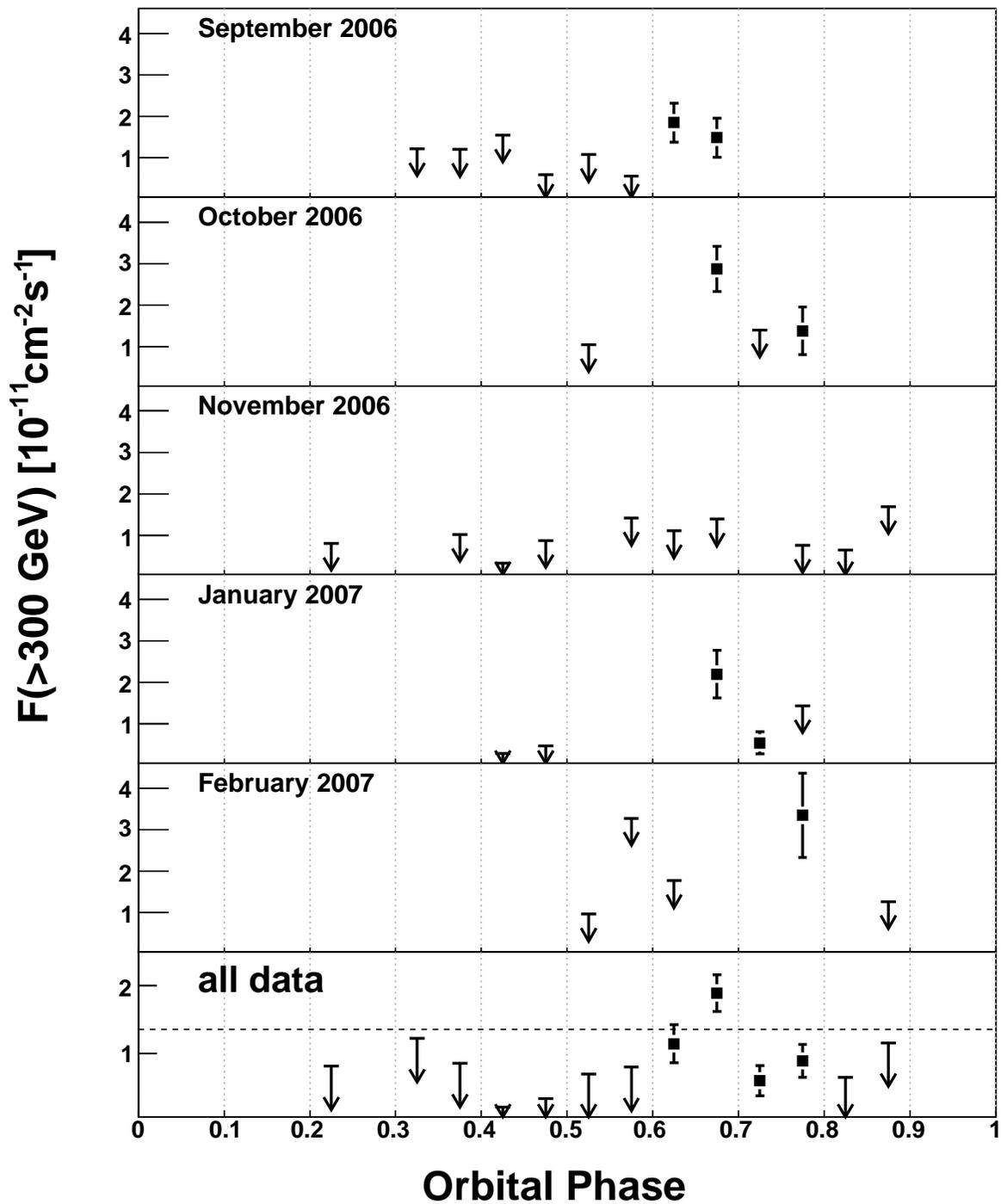}
\caption{\label{fig:fluxVsPhase}
Average fluxes or upper flux limits per orbital phase bin for $\gamma$-rays with energies above
300 GeV from the direction of \mbox{LS I +61 303} as a function of orbital phase.
The bottom panel shows the results averaged over the whole dataset, the 
upper panels show the results for individual orbits.
Upper flux limits (95\% probability, \citet{Helene-1983}) are shown for data points with significances
less than $2\sigma$ (significance calculation after \citet{Li-1983}, equation 17).
A flux corresponding to 10\% of the flux of $\gamma$-rays from the Crab Nebula is indicated by the dashed 
horizontal line in the bottom panel.
}
\end{figure}

The dependency of integral fluxes
on orbital phase
(Figure \ref{fig:fluxVsPhase}) shows that significant fluxes above $2\sigma$
are only detected during orbital phases close to apastron.
Peak fluxes measured at phases between 0.6 and 0.8
correspond to 10-20\% of the flux of the Crab Nebula,
but appear to vary from one orbital cycle to the next.
Upper limits in a range corresponding to 3-10\% of the flux of the Crab Nebula have been
derived for phases 0.3 to 0.6 and around 0.9.
The upper limit calculation assumes a similar shape for the energy spectra of 
TeV $\gamma$-rays during all phases.
This may not be correct, as the phase-dependent shape
of the energy spectra of the 
\mbox{X-ray} binary LS 5039 shows \citep{Aharonian-2006b}.
Furthermore, the combination of data from different orbital cycles in Figure \ref{fig:fluxVsPhase} 
(bottom) assumes implicitly that the position of the emission maximum in orbital phase and
its flux do not vary from one cycle to the next.
A constant maximum $\gamma$-ray flux is not necessarily expected;
\mbox{LS I +61 303} shows variable X-ray fluxes over different orbits (e.g.~Chernyakova et al. 2006).
The VERITAS measurement
can neither confirm nor refute this at this stage.
The same periodicity for radio and $\gamma$-ray data was assumed during this analysis (P=26.4960 days).
This has been tested using Lomb-Scargle statistics \citep{Scargle-1982}, but
large gaps in data taking due to bright moon periods and 
the periodicity of the observation cycles
do not allow definite conclusions.

\section{Discussion and Conclusions}

\mbox{LS I +61 303} has been detected in $\gamma$-rays
at phases around 0.6 to 0.8 only, when the distance between the two
objects in the binary system is largest.
This indicates a strong dependence of particle acceleration and energy loss mechanisms
on the relative positions.

It is suggested that TeV $\gamma$-rays are produced in \mbox{LS I +61 303}  
in leptonic (inverse Compton scattering of low-energy
photons from the stellar companion or pair cascades,
see e.g.~\citet{Bednarek-2006, Gupta-2006}) 
and/or hadronic interactions (e.g.~\citet{Romero-2005}).
Both production mechanisms require a population of high-energy particles
and a sufficiently dense photon field or target material.
This is provided abundantly in \mbox{LS I +61 303} for all orbital phases
by stellar photons and the wind of the companion star.
Target density is highest around periastron, which
favors a maximum in $\gamma$-ray emission 
when the compact object passes close to 
the stellar companion.
The maximum electron energy available for $\gamma$-ray production
is, on the other hand, restricted
by cooling through inverse Compton 
and synchrotron emission; both of these effects can be strongest around periastron.
In addition, $\gamma$-rays emitted too close to the massive star
suffer from photon-photon absorption in the dense
stellar radiation field.
Significant inhomogeneities in the stellar wind can further complicate
the process of $\gamma$-ray production \citep{Neronov-2007, Romero-2007b}.

The unknown nature of the compact object (neutron star or black hole) allows
two very different scenarios to provide the necessary power for 
the production of VHE $\gamma$-rays.
In the microquasar model, charged particles (electrons or hadrons) are
accelerated in an accretion-driven relativistic jet
\citep{Taylor-1984,Mirabel-1994}.
Acceleration can take place all along the jet or in shocks created in the jet termination 
zone \citep{Heinz-2002}.
The efficiency of particle acceleration and $\gamma$-ray production varies over the orbit due
to changes in target density, accretion rate, and magnetic field strength.
The second scenario assumes that the compact object is a pulsar.
Particles are accelerated in the shock created 
by the collision of the expanding
pulsar wind with the equatorial disk or wind of the companion star
\citep{Maraschi-1981}, similar to the binary system \mbox{PSR B1259-63}.
The distance of the pulsar wind termination shock 
changes with orbital motion \citep{Dubus-2006b}.
Electron cooling through synchrotron emission is in this model lowest around apastron, when the
shock is furthest away from the pulsar and therefore the magnetic field is weakest.
This results, under the assumption that enough target photons are available for inverse Compton scattering,
in the observed maximum of the $\gamma$-ray emission at apastron.
More measurements, especially contemporaneous multiwavelength observations are necessary
for constraining tests of the available models.

The measured integral flux
of $\gamma$-rays above 300 GeV in the phase interval 0.6 to 0.8
corresponds to an isotropic luminosity of roughly $10^{34}$ ergs/s at 2 kpc.
The necessary power could either be provided by a 
pulsar with spindown luminosity of $~\sim10^{36}$ erg/s or by accretion, which has
been estimated to be in the range of several $10^{37}$ erg/s
\citep{Romero-2007}.
Both scenarios imply conversion factors to $\gamma$-ray emission below 1\%.

%
%
%
%
%
%
If we compare the average TeV flux with the average GeV flux measured by EGRET \citep{Kniffen-1997},
we find a GeV/TeV photon flux ratio of $10^5$ which is consistent with that produced by a power-law with photon
index 2.4, similar to the photon indexes measured in the two bands.
This result should be treated with caution since the source
is known to exhibit long-term
variability \citep{Gregory-1989} and the measurements are separated by more than 10 years.
However, it may suggest that a single power-law could suffice to describe the GeV/TeV spectrum
- a possibility that can be tested with future simultaneous GeV/TeV observations.

In conclusion, new stereoscopic observations with VERITAS have confirmed very-high-energy 
$\gamma$-ray emission from \mbox{LS I +61 303}.
The source is detected only during
orbital phases 0.6 to 0.8, which is close to apastron.
This suggests variability connected to the orbital movement,
but a definite conclusion has to wait for further measurements.
Future observations with
the much more sensitive full VERITAS array will allow us to study the
VHE emission during the whole orbital cycle.

\acknowledgments

This research is supported by grants from the U.S.~Department of Energy, the National Science Foundation,
the Smithsonian Institution, by NSERC in Canada, by Science Foundation Ireland and by PPARC in the UK.
We acknowledge the excellent work of the technical support staff at the FLWO and 
the collaborating institutions in the construction and operation of the instrument.


\clearpage

\begin{deluxetable}{rccccc}
\centering
\tablecolumns{6}
\tablewidth{0pt}
\tablecaption{
Details of the VERITAS observations of \mbox{LS I +61 303}.
\label{tab:observations}}
\tablehead{
\colhead{Month} &
\colhead{N$_{\mathrm{Tel}}$\tablenotemark{a}} &
\colhead{Wobble} &
\colhead{Elevation} &
\colhead{Obs.} &
\colhead{Orbital} \\
\colhead{} &
\colhead{} &
\colhead{offset} &
\colhead{range} &
\colhead{time\tablenotemark{b}} &
\colhead{phase} \\
\colhead{} &
\colhead{} &
\colhead{} &
\colhead{} &
\colhead{ [h] } &
\colhead{interval}
}
\startdata
09/2006  & 2 & 0.3$^{\mathrm{o}}$ & 53-61$^{\mathrm{o}}$ & 12.5  & 0.31-0.69 \\
10/2006  & 2 & 0.3$^{\mathrm{o}}$ & 61-61$^{\mathrm{o}}$ & 7.2   & 0.51-0.78 \\
11/2006  & 2 & 0.3$^{\mathrm{o}}$ & 54-61$^{\mathrm{o}}$ & 12.8  & 0.23-0.87 \\
01/2007  & 3 & 0.5$^{\mathrm{o}}$ & 55-61$^{\mathrm{o}}$ & 9.9   & 0.41-0.76 \\
02/2007  & 3 & 0.5$^{\mathrm{o}}$ & 52-57$^{\mathrm{o}}$ & 3.5   & 0.51-0.89 \\
\hline
Total    &    &                   & 52-61$^{\mathrm{o}}$ & 45.9  & 0.23-0.89 
\enddata 
\tablenotetext{a}{Number of available telescopes}
\tablenotetext{b}{Dead-time-corrected observation time}
\end{deluxetable}

\clearpage

\end{document}